\documentclass[11pt]{article}
\usepackage[a4paper, total={6.5in, 9in}]{geometry}
\usepackage{amsmath}
\usepackage{amsfonts}
\usepackage{mathtools}
\usepackage{hyperref}
\usepackage{breqn} %for breaking equation in two lines
\usepackage{bm}
\usepackage{graphicx}
\usepackage{epstopdf}
\usepackage{subcaption}
\usepackage{authblk}
\usepackage{color,soul}
\usepackage{tikz-cd} 

\DeclarePairedDelimiterX{\norm}[1]{\lVert}{\rVert}{#1}
\def\equationautorefname~#1\null{Equation (#1)\null}

\title{A Weyl geometric model for thermo-mechanics of solids with metrical defects}
\author[1]{Bensingh Dhas\thanks{bensinghdhas@iisc.ac.in}}
\author[3]{Arun R Srinivasa \thanks{asrinivasa@tamu.edu}}
\author[1,2]{Debasish Roy\thanks{royd@iisc.ac.in}}
\affil[1]{Centre of Excellence in Advanced Mechanics of Materials, Indian
Institute of Science, Bangalore 560012, India }
\affil[2]{Computational Mechanics Lab, Department of Civil Engineering, Indian
Institute of Science, Bangalore 560012, India}
\affil[3]{Department of Mechanical
Engineering, Texas A{\&}M University, College Station, Texas 77843-3123}

\date{}

\newcommand{\bs}{\boldsymbol}
\newcommand{\tb}{\textbf}
\begin{document}
\maketitle

\begin{abstract} 
    This work seeks a rational route to large-deformation, thermo-mechanical
    modeling of solid continua with metrical defects.  It assumes the
    geometries of reference and deformed configurations to be of the Weyl type
    and introduces the Weyl one-form -- an additional set of geometrically
    transparent degrees of freedom that determine ratios of lengths in different
    tangent spaces.  The Weyl one-form prevents the metric from being compatible
    with the connection and enables exploitation of the incompatibility for
    characterizing metrical defects in the body.  When such a body undergoes
    temperature changes, additional incompatibilities appear and interact with
    the defects. This interaction is modeled using the Weyl transform, which
    keeps the Weyl connection invariant whilst changing the non-metricity of the
    configuration. 
    An immediate consequence of the Weyl connection
    is that the critical points of the stored energy are shifted. We exploit
    this feature to represent the residual stresses. In order to relate stress and strain in our non-Euclidean
    setting, use is made of the Doyle-Ericksen formula, which is interpreted as a
    relation between the intrinsic geometry of the body and the stresses
    developed. Thus the Cauchy stress is conjugate to the Weyl transformed
    metric tensor of the deformed configuration. The evolution
    equation for the Weyl one-form is consistent with the two laws of
    thermodynamics.  Our temperature evolution equation, which couples
    temperature, deformation and Weyl one-form, follows from the first law of
    thermodynamics. Using the model, the self-stress generated by a
    point defect is calculated and compared with the linear elastic solutions.
    We also obtain conditions on the defect distribution (Weyl
    one-form) that render a thermo-mechanical deformation stress-free. Using
    this condition, we compute specific stress-free deformation profiles for
    a class of prescribed temperature changes.  
    \end{abstract} 

\section{Introduction}
The inelastic response of a material body is brought about by anomalies in their
internal structure -- what is loosely referred to as 'geometric frustration'. In
materials with structural order (crystalline materials, to wit), these anomalies
are identified with dislocations and disinclinations.  Anomalies in structured
solids have been studied using tools from differential geometry over a long
time. The work by Bilby \textit{et} \textit{al.} \cite{bilby1955} is a classical
example where geometric tools were employed to study dislocations in a
crystalline solid.  They addressed the technologically important problem of
describing dislocations using a continuous field and constructed a geometric
space with an asymmetric affine connection whose torsion was related to the
density of dislocations;  see \cite{kroner1990} for a differential geometric
description.  To generalize such a description of defects, Wang \cite{wang1967}
introduced the notion of a material connection, which was characterized using
the material uniformity field. The parallel transport of vectors between
different tangent spaces was aided by this material uniformity field. Invariants
of the material connection describing the intrinsic geometry of the material
body (torsion, curvature etc.) were related to the density of defects.

Inelasticity at the continuum scale, on the other hand, is often modelled with
internal variables. Phenomena such as thermo-elasticity, visco-elasticity,
visco-plasticity, growth, damage etc. fall within the scope of such a
description; a brief introduction to a class of these theories may be found in
Gurtin \textit{et} \textit{al.}\cite{gurtin2010}. These theories introduce
scalar, vector or tensor internal variables, with little or no geometric
(kinematic) interpretation. As a consequence, the deformation field is often
coupled with these internal variables somewhat arbitrarily. Moreover, the idea
of an intermediate configuration, frequently used with these theories, is hard
to appreciate from both physical and mathematical standpoints.  The
elasto-plastic decomposition, used in finite deformation plasticity, is an
example that exploits the notion of an intermediate configuration via a
multiplicative decomposition of the deformation gradient; see \cite{sadik2015}
for a brief historical review. As the geometry of the intermediate configuration
is opaque, its introduction also corrupts notions of differentiation for vector
and tensor fields.  On the whole, the clarity offered by geometrically motivated
micro-mechanical theories of Kondo, Bilby and others is often lost in such a
description of inelasticity at the continuum scale.

The assumption of a fixed relaxed configuration has been questioned before by
Eckart \cite{eckart1948} . He used an anelasticity tensor to characterize the
anelastically deformed solid; this tensor was determined based on an evolution
equation for the metric tensor.More recently  Rajagopal and Srinivasa \cite{rajagopal2004thermomechanics}  used a notion of evolving natural configurations to describe inelastic response of a variety of materials. On similar lines, a geometrically founded
continuum thermo-mechanical theory was proposed by Stojanovitch and co-workers
\cite{stojanovitch1969}. It was based on the decomposition of the deformation
gradient into elastic and thermal parts.  The individual parts were assumed to
be non-integrable, even as the total deformation gradient remained integrable.
A frame field having non-zero anholonomy was defined using the thermal part of
the deformation gradient. The Weitzenbok connection was constructed by demanding
the frame fields to be parallel.  It should be mentioned that the connection
utilized by Stojanovitch and co-workers was the same used by Bilby and others to
describe a dislocated solid. This modeling approach does not affect the metric
on the body manifold. This implies that the distance between material points do
not change as the temperature of the body is varied from the reference
temperature; hence no strains are introduced. Because of this property, the
Weitzenbok connection does not appear to offer an attractive route for modelling
thermo-elasticity.  Ozakin and Yavari \cite{ozakin2010} have proposed an
alternate geometric framework for thermo-elastic deformations. Their work is
based on a hypothesized material or relaxed manifold whose geometry is
non-Euclidean. In such a configuration, a thermally strained body is stress
free.  They implement this hypothesis by allowing the metric tensor of the
material manifold temperature dependent. Deformation now becomes a map between
the material and spatial configurations. The difference in the geometry of
spatial and material manifolds leads to stresses in the former. Using this
setting, they obtain deformation and temperature fields which do not produce
stresses. The condition for a temperature change to be stress free is worked out
as the vanishing of the curvature tensor of the material manifold.  However, the
free energy is not a function of the curvature of the material manifold. A major
problem with the worldview of Ozakin and Yavari is posed by the material
manifold -- the premise that the material manifold is stress-free cannot be
verified since the notion of stress for the material manifold is undefined.
Moreover, having a configuration which is virtual (i.e.  physically
unrealizable) only increases the opacity of the theory. The geometric
description of thermo-elastic bodies discussed in \cite{ozakin2010} has been
extended to include time evolution by Sadik and Yavari \cite{sadik2017}.

Considerable effort has gone in the modelling of point defects, biological
growth and thermal strains through an internal variable perspective. Models
belonging to this genre try to capture the local zones of expansion or
contraction using internal variables. Cowin and Nunziato \cite{cowin1983}
introduced the void volume fraction as an additional kinematic variable which
followed a second order evolution rule. Based on this, they demonstrated that a
porous material could support two kinds of waves: one determined by its elastic
properties and the other by properties related to porosity. Similarly,
Garikipati \textit{et} \textit{al.} \cite{garikipati2006} developed a continuum
formulation for point defects within the framework of linear elasticity.  Point
defects were understood as centers of expansion or contraction. A formation
volumetric tensor was introduced to characterize point defects;  the dipole
tensor conjugate to the formation volumetric tensor described the forces
required to keep the point defect in mechanical equilibrium. The linear nature
of the theory permitted the use of Green's functions to determine the stresses
due to the point defects.  In a more recent work, Moshe \textit{et} \textit{al.}
\cite{moshe2015} has developed a metric description for defects in amorphous
solids.  Here, the absence of long range order in the material makes it
impossible to define Burgers' vector in terms of the torsion tensor.  However,
using the Levi-Civita parallel transport, the authors arrive at a notion for
Burgers' vector. They also show that, in a two dimensional setting, a conformal
change of the metric tensor is sufficient to represent dislocation- and
disinclination-like defects in amorphous solids. The question of
restructurability of geometrically frustrated solids is addressed by Zurlo and
Truskinovsky \cite{zurlo2017}. They develop a surface deposition protocol to
additively manufacture these residually stressed bodies. Yavari and Goriely
\cite{yavari2012} formulate a geometric theory for point defects. They postulate
that the stress-free configuration of a body with point defects may be
identified with a Weyl manifold. The density of defects in the body is defined
as the deviation of the volume form of the material manifold from the Euclidean
volume form.  Using this, they compute the stresses generated by a shrink-fit in
the finite deformation setting. To represent the shrink fit, the volume form is
assumed to be discontinuous across the shrink fit surface; this is however not
in conformity with the structure of a differentiable manifold. They also employ
the usual equilibrium equations without explicating on its variational
structure, even though the divergence of stress depends on the Weyl connection.
Moreover, with the focus being only on analytical solutions, the constitutive
framework adopted by Yavari and Goriely remains somewhat simplistic. For
instance, no information on distant curvature was made use of in the model.

Though an important problem, the modelling of the mechanical response of solids
with metrical defects and temperature change is inadequately addressed in the
literature. By metrical defects we mean the anomalies in the material body,
which modify the local notion of length. Point defects and growth are examples
of metric anomalies. A key component in building such a theory is a geometric
space that consistently incorporates the modified notion of length due to
defects and temperature change. This is what we set out to pursue, using a Weyl
geometric setting. Weyl manifolds are geometric spaces where the metric is
incompatible with the connection. This incompatibility has previously been
exploited to describe point defects \cite{yavari2012, roychowdhury2017}; however
there are considerable differences in our present worldview.  The Weyl transform
which keeps the connection invariant but alters the metric by a positive factor
is now thought of as a modification introduced in the configuration of the body
due to temperature change. Identifying the reference and deformed configurations
as Weyl manifolds make it possible to represent defects and temperature changes
in a consistent manner. This modeling perspective also has an important
advantage: it completely avoids the mysterious intermediate configuration. The
inelastic response of the body arising out of the defect distribution is buried
within the connection associated with the configuration. An important
consequence of having a connection different form the flat Levi-Civita is that
the critical points of the stored energy function are non-trivially modified.
This modification may be exploited to represent configurations which are
residually stressed.  This fact, to the best of our knowledge, remains
unexplored in the literature on defect mechanics. The description has a parallel
with the idea of 'pseudo-'forces encountered in particle mechanics; these forces
arise only when the components of the connection are non-trivial. This approach
is used to predict the stresses created by a diffused shrink fit.  The
prediction of the radial stress by our methodology is found to be in line with
that of the linear elastic solution; however the hoop stress is found to be
quite different. The deviation in the hoop stress is essentially due to the
diffused representation of the shrink fit. Further, the model is applied to
arrive at a condition for a body to be in a state of zero stress under a
prescribed temperature change. Using this condition we also recover a well know
stress free configurations known from linear elasticity.

This article is organized into eight sections and an appendix. Section 2 gives a
brief introduction to a Weyl manifold and discusses its key invariants.
Kinematics of a body with metrical defects are discussed in section 3. Kinetics
required to describe the dynamics of the defective body are considered in
section 4; an important aspect being the adaptation of the classical
Doyle-Ericksen formula within the Weylian setting. The equilibrium equation of
the Weyl one-form is also derived in this section. In section 5, restrictions
imposed by the laws of thermodynamics on the constitutive relations are
discussed. Restricting the theory only to local equilibrium considerations, self
stresses generated due to a point defect are discussed in section 6. Conditions
for a configuration to be stress free state is discussed in section 7. Section 8
concludes the article with a few observations and comments on further possible
applications of the present model. The appendix briefly discusses the numerical
procedure used to solve the diffused shrink fit problem.

\section{Weyl geometry}
The defining hypothesis of a Riemannian manifold is that the positive definite
metric is preserved under parallel transport. H Weyl \cite{weyl1921electricity,
weyl1918gravitation}, in an effort to include electromagnetism within the
framework of general relativity, made the following hypothesis: the parallel
transport of the metric along a curve is proportional to the metric itself. In
other words, the ratios of length between different tangent spaces also need to
be parallel transported.  This hypothesis introduces an additional degree of
freedom which generalizes the Riemannian geometry by an independent scale at each
tangent space of the manifold. Even though Weyl's relativity did not succeed in
unifying general relativity and electro-magnetism, the resulting mathematical tool was
used to describe other physical phenomena (for recent developments in Weyl
relativity, see \cite{barcelo2017}). We now briefly review the construction of the
Weyl geometry. Let $\mathcal{M}$ be an $n$ dimensional differentiable manifold
with a Riemannian metric $\tb{g}$ and $\gamma$ be a curve in $\mathcal{M}$ given
by, $\gamma:(0,1) \rightarrow \mathcal{M}$. Using coordinates, the curve may be
given by $(x^1(t),..,x^n(t)) \in \mathcal{M}$, $t\in (0,1)$.  The infinitesimal
version of Weyl's hypothesis may be written as,
\begin{equation}
\frac{d}{d t}g(V,W)=\phi(\gamma'(t))g(V,W)
\label{eq:weylHypothesis}
\end{equation}
$V,W \in T_x \mathcal{M}$, where $T_x \mathcal{M}$ denotes the tangent space at
$ x \in \mathcal{M}$ and $\gamma'(t)$ is the vector field tangent to $\gamma$ at
$(x^1(t),..,x^n(t))$. The one-form $\phi$ is introduced to describe the scale
degrees of freedom at each tangent space. Note that as $\phi$ is set
to zero, Eq.  \ref{eq:weylHypothesis} reduces to the defining hypothesis of
Riemannian geometry.  We now formally define a Weylian manifold to be the triplet
 $(\mathcal{M}, \tb{g}, \phi)$, consisting of a differential manifold $\mathcal{M}$, a
Riemannian metric $\tb{g}$ and a one-form $\phi$. On integrating Eq.
\ref{eq:weylHypothesis} along the curve $\gamma$, we arrive at a relation
between inner-products at $T_{x(\gamma(0))}\mathcal{M}$ and
$T_{x(\gamma(t_1))}\mathcal{M}$ given by,
%We compute the
%connection induced by the Weyl condition given in Eq. \ref{eq:weylHypothesis}.
\begin{equation}
    g(V(t),W(t))=g(V(0),W(0))e^{\int_0^{t_1} \phi(\gamma'(t)) dt}.
	\label{eq:integralWeyl}
\end{equation}
We denote the length of a vector $V$ by $l=g(V,V)$. To obtain the variation
of $l$ along $C$, we set $V=W$ in Eq. \ref{eq:integralWeyl},  leading to the
following relationship,
\begin{equation}
    l(s)=l(0)e^{\int_0^{t_1} \phi(\gamma'(t)) dt}
\end{equation}
If we choose the curve to be closed, i.e. $\gamma(0)=\gamma(1)$, the transport of the inner-product
given in Eq. \ref{eq:integralWeyl} becomes,
\begin{equation}
	g(V(1),W(1))=g(V(0),W(0))e^{\oint \phi(\gamma'(t)) dt}
	\label{eq:loopWeyl}
\end{equation}
Using Stokes' theorem for the line integral, the equation above may be written as,
\begin{equation}
    \int_{\Omega} d\phi=\oint_{\gamma}\phi
\end{equation}
where, $\Omega$ is the area enclosed by the closed curve $\gamma$.
If the line integral in Eq. \ref{eq:loopWeyl} vanishes for any closed curve $\gamma$,
then we have $d\phi=0$. Now using Poincar\'es' lemma, we have an exact Weyl one-form
$\phi$:
\begin{equation}
\oint \phi(\gamma'(t)) dt=0 \implies \phi=d f
    \label{eq:integralityCondition}
\end{equation}
where $f$ is a scalar valued function and $df$ denotes the differential
of $f$. Thus integrability of the Weyl one-form ensures that the parallel
transport of the inner-product is path independent. We also define an integrable
Weyl manifold as one whose metric has a path independent parallel transport.
Formally, an integrable Weyl manifold is denoted by $(\mathcal{M},\tb{g}, d f)$.
If $\phi$ does not satisfy the integrability condition given in Eq.
\ref{eq:integralityCondition}, then the parallel transport is path dependent; we call such
a manifold non-integrable Weyl.

\subsection{Weyl connection}
An affine connection is an additional structure defined on a smooth manifold,
using which one can differentiate vector and tensor fields. It also defines a
notion of parallel transport on a tangent bundle. On a differentiable manifold,
one can define infinitely many connections. However, we consider the unique torsion
free connection which is natural to Weyl's hypothesis. In terms of covariant
derivatives, Weyl's hypothesis may be written as,
\begin{equation}
\nabla \tb{g} =\phi \tb{g} 
\label{eq:nonmeticity}
\end{equation}
where, $\nabla(.)$ is the covariant derivative relative to the Weyl connection.
In component form, the equation may be written as,
\begin{equation}
g_{ij;k}=\phi_kg_{ij}
\label{eq:WeylCovariant}
\end{equation}
We introduce a new tensor field $\tb{Q}:=\phi \otimes \tb{g}$, called the
non-metricity tensor, which describes the incompatibility of the Weyl connection
with the metric.  In the equation above, $g_{ij;k}$ denotes the  covariant
derivative of the metric tensor. With respect to an arbitrary connection, $g_{ij;k}$ may be written as,
\begin{equation}
g_{ij;k}=\partial_k g_{ij}-\Gamma_{ki}^l g_{lj}-\Gamma_{kj}^l g_{il}
\label{eq:covDiffMetric}
\end{equation}
Here $\Gamma_{ij}^k$ denotes the Christoffel symbol of the
second kind.
The assumption that the connection is torsion free implies a
symmetry condition on the lower two indices of the Christoffel symbol,
$\Gamma^k_{ij}=\Gamma^k_{ji}$.  Using Eq. \ref{eq:covDiffMetric} in Eq.
\ref{eq:WeylCovariant}, along with the torsion free assumption leads to the
following expression for the coefficients of the Weyl connection.
\begin{equation}
\Gamma_{mi}^l=\frac{1}{2}g^{kl}\left(\partial_m g_{ik}+\partial_i g_{km}-\partial_k g_{mi}\right)
-\frac{1}{2}(\phi_m \delta_i^l +\phi_i \delta_m^l - \phi_j g_{mi}g^{jl})
\label{eq:weylConnection}
\end{equation}
If the Weyl one-form is integrable then we have,
\begin{equation}
\Gamma_{mi}^l=\frac{1}{2}g^{kl}\left(\partial_m g_{ik}+\partial_i g_{km}-\partial_k g_{mi}\right)
-\frac{1}{2}(\partial_m f \delta_i^l +\partial_i f  \delta_m^l - \partial_j f  g_{mi}g^{jl})
\label{eq:integWeylConnection}
\end{equation}
Note that for any Weyl manifold (integrable or non-integrable), the connection
may be written as the sum of the Levi-Civita connection and a (1,2) tensor. This
(1,2) tensor is determined by the metric and the Weyl one-form. Another
important property of a Weyl connection is the invariance under Weyl
transformation. A Weyl transformation is defined by,
\begin{equation}
    g_{ij}\rightarrow e^s g_{ij};\;\;\; \phi \rightarrow \phi+ds
    \label{eq:weylTransform}
\end{equation}
where, $s$ is a real valued function. One may immediately verify the
invariance of the connection by substituting Eq.  \ref{eq:weylTransform} in Eq.
\ref{eq:weylConnection}. However, if we define $\bar{g}_{ij}:=e^sg_{ij}$, 
$\bar{\phi}:=\phi+ds$ and substitute them in Eq.  \ref{eq:covDiffMetric}, we
arrive at the following,
\begin{equation}
    \bar{g}_{ij;k}=\bar{\phi}_k \bar{g}_{ij}
\end{equation}

The invariance of the Weyl connection under Weyl transformation is a cornerstone
in our thermo-mechanical theory; section \ref{sec:kinematics} discusses this aspect in detail.

\subsection{Equivalance of an intergable Weyl manifold and Riemannina manifold}
The notion of an integrable Weyl manifold was established at the beginning of this
section. We now discuss its equivalence with a Riemannian  manifold.  This
equivalence is established by showing that Weyl's connection boils down to the
Levi-Civita connection when the Weyl one-form is exact. For the sake of
completeness we record the Levi-Civita connection induced on a Riemannian manifold
by the metric $\tb{g}$. The Christoffel symbols associated with the Levi-Civita
connection are given by,
\begin{equation}
\Gamma_{mi}^l=\frac{1}{2}g^{kl}\left(\partial_m g_{ik}+\partial_i g_{km}-\partial_k g_{mi}\right)
\end{equation}
By setting the Riemannian metric as $e^f \tb{g}$ and computing the
Christoffel symbols of the associated Levi-Civita connection, we are led to the
connection coefficients of an integrable Weyl connection given in Eq.
\ref{eq:integWeylConnection}. This result implies that an
integrable Weyl manifold is a Riemannian manifold with a modified metric.
%, however by suitably defining
%the kinetics on a integrable Weyl manifold, one can show that it can model a
%larger class of phenomena.

\subsection{Invariants of a Weyl connection}
On a Weyl manifold, the presence of a metric permits us to define the
arc length for any curve on the manifold. This is similar to the case with Riemannian
manifolds. Let $\gamma$ be a curve on $\mathcal{M}$; the expression for the 
arc-length of $\gamma$ is given by,
\begin{equation}
    l=\int_{\gamma} e^{\frac{s}{2}} \sqrt{\phi\left(\frac{d \gamma}{dt}\right)\tb{g}\left(\frac{d \gamma}{dt},
    \frac{d \gamma}{dt}\right)}dt
\end{equation}
$\tb{g}(.,.)$ is the Riemannian metric associated with a Weyl manifold,
$\frac{d\gamma}{dt}$ is the vector field tangent to $\gamma$ and $s$ is the
variable of the Weyl transformation.  Note that $s$ weights the metric at each point on
the curve. We choose the following three-form as the volume form,
\begin{equation}
    d\tb{V}=e^{\frac{ns}{2}}\sqrt{g}dx^1\wedge...\wedge dx^n
\end{equation}
where $n$ is the dimension of the manifold and $g$ the determinant of the
Riemannian metric. Note that the volume-form is not compatible with the
Weyl connection. Apart from non-metricity, the Weyl connection may also have
curvature.  As with a Riemannian manifold, the curvature operator is defined as
the non-commutativity of second covariant derivatives.
\begin{equation}
\tb{R}(X,Y)(Z)=\nabla_X \nabla_Y Z- \nabla_Y \nabla_X Z-\nabla_{[X,Y]} Z
%;\;\;\;
%X,Y,Z \in \text{Vec}(\mathcal{M})
\end{equation}
In the equation above, $X,Y$ and $Z$ are vector fields on $\mathcal{M}$. The
curvature operator is linear in all its input arguments; it is also
anti-symmetric in $X$ and $Y$.  If one writes the curvature operator in a
coordinate (holonomic) basis, the last term in the above equation vanishes. The
components of the curvature tensor are given by,
\begin{equation}
    R^l{}_{ikj} = \partial_k\Gamma^l{}_{ji} -
    \partial_j\Gamma^l{}_{ki} +
    \Gamma^l{}_{ka}\Gamma^a{}_{ji} -
    \Gamma^l{}_{ja}\Gamma^a{}_{ki}
\end{equation}
The Ricci curvature is given by,
\begin{equation}
R_{ij}=R^k_{ikj}
\end{equation}
In contrast to the Riemannian case, the Ricci tensor of a Weyl manifold is 
unsymmetric and thus may be  decomposed into its symmetric and 
antisymmetric parts.
\begin{equation}
R_{ij}=\hat{R}_{ij}+K_{ij}
\end{equation}
$\hat{R}_{ij}$ and $K_{ij}$ respectively denote the symmetric and antisymmetric parts of the Ricci tensor. The antisymmetric tensor $\tb{K}$ is sometimes called the 
distant curvature. It is given by the exterior derivative of the Weyl one-form.
\begin{equation}
\tb{K}=d \phi
\end{equation}
For an integrable Weyl manifold, the distant curvature vanishes making the Ricci tensor 
symmetric; this follows from the identity $dd=0$.  Using the symmetric
part of the Ricci tensor, the following scalar curvature
can be defined,
\begin{equation}
R=g^{ij}R_{ij}
\end{equation}
Note that $R$ does not depend on the distant curvature. The
scalar measure extracted form the anti-symmetric distant curvature is 
\begin{equation}
K=g^{ik} g^{jl}K_{ij}K_{kl}
\end{equation}
Using the machinery of Weyl manifolds discussed so far, we now proceed to develop
a thermo-mechanical theory for a solid body with metrical defects.

\section{Kinematics}
\label{sec:kinematics}
We assume the reference and deformed configurations to be Weylian manifolds.
These configurations are respectively denoted by the triplets $(\mathcal{B},
\tb{G},\xi)$ and $(\mathcal{S},\tb{g},\phi)$, where $\mathcal{B}$ and
$\mathcal{S}$ are smooth three-manifolds.  The metric tensors of the reference
and deformed configurations are denoted by $\tb{g}$ and $\tb{G}$ and the
associated Weyl one-forms by $\eta$ and $\phi$ respectively. We introduce an
additional kinematic variable in the deformed configuration called the Weyl
scaling, which is denoted by $s$. $s$ is identified as the variable of Weyl
transformation discussed in the previous section. The Weyl transformation
modifies the deformed metric by scaling it by a positive scalar and changing the
one-form by an exact one-form $ds$. In this work, we assume that Weyl
transformation models the thermal strain introduced in the body due to
temperature change. Since the Weyl connection is invariant under Weyl
transformation, the Weyl scaling does not modify the connection. However the
metric and the non-metricity are modified due to temperature change. This
modification represents the change in the strain field and defect distribution
in the material body.  The Weyl transformed metric and the one-form of the
deformed configuration are denoted by $\bar{\tb{g}}:=e^s\tb{g}$ and
$\bar{\phi}=\phi+ds$ respectively. We relate $s$ to the temperature field of the
deformed configuration through the following relationship,
\begin{equation}
    s=\alpha (T-T_0).
    \label{eq:thermalExpansion}
\end{equation}
The equation above is a constitutive relation chosen for simplicity, although
other relationships are possible. The variables $\alpha$, $T$ and $T_0$
respectively denote the co-efficient of thermal expansion, temperature field and
reference temperature of the body.  Throughout this work, we assume the metric
tensors $\tb{G}$ and $\tb{g}$ as Euclidean. The deformation map relating the
reference and deformed configurations is given by, 
\begin{equation}
    \bs{\varphi}:\mathcal{B}\rightarrow \mathcal{S}.
\end{equation}
The derivative of the deformation map or the tangent map is denoted by $\tb{F}$;
it maps tangent vectors from the reference configuration to those of the
deformed configuration. If we denote the co-ordinates of the reference and
deformed configurations by $(X^1,...,X^3)$ and $(x^1,...,x^3)$, then the
deformation gradient can be expressed as,
\begin{equation}
\frac{\partial }{\partial x^i}=F_i^I\frac{\partial}{\partial X^I}
\end{equation}
where, $\frac{\partial}{\partial X^I}$ and $\frac{\partial}{\partial x^i}$ are
the coordinate bases for $T_X\mathcal{B}$ and $T_{\varphi(X)}\mathcal{S}$
respectively. These basis vectors may be identified with vectors tangent to the
coordinate lines at each point of the manifold. Whenever the deformed
configuration has a metric tensor, the right Cauchy-Green deformation tensor
$\tb{C}:T_X\mathcal{B} \times T_X\mathcal{B}\rightarrow \mathbb{R}$ can be
defined by pulling back the metric tensor of $\mathcal{S}$ to $\mathcal{B}$:
\begin{equation}
    \tb{C}=\bs{\varphi}^*\tb{g}.
\end{equation}
In this equation, $\bs{\varphi}^*(.)$ denotes the pull-back map. The
covariant components of $\tb{C}$ are thus given by, 
\begin{equation}
    C_{IJ}=F^i_Ig_{ij}F^j_J
    \label{eq:cauchGreenUsual}
\end{equation}
As the temperature of the body changes from the reference, the
metric and the Weyl one-form of the deformed configuration change via the Weyl
transformation. In the presence of thermal strains, the Cauchy-Green deformation
tensor denoted by $\bar{\tb{C}}$ is given by,
\begin{align}
    \nonumber
    \bar{C}_{IJ}&=F^i_I\bar{g}_{ij}F^j_J\\
    &=e^sF^i_Ig_{ij}F^j_J
    \label{eq:cauchGreenScaled}
\end{align}
From now on, we will use the expression in Eq. \ref{eq:cauchGreenScaled} as the
definition of the right Cauchy-Green deformation tensor. Note that Eq.
\ref{eq:cauchGreenScaled} reduces to the usual definition of the right
Cauchy-Green deformation tensor when $s$ is set to zero. This pertains to the
case of no temperature change in the body. Components of the Green-Lagrangian
strain in the presence of temperature change are given by,
\begin{equation}
    E_{IJ}=\frac{1}{2}(\bar{C}_{IJ}-G_{IJ})
\end{equation}
As a consequence of Weyl transformation, the principal stretches
are scaled by $\exp\left(\frac{s}{2}\right)$. If $\lambda_i$ are the
principal stretches of $\tb{C}$, it follows that,
\begin{equation}
    \bar{\lambda}_i^2=\exp(s)\lambda_i^2
\end{equation}
$\bar{\lambda}_i$ are the principal stretches of $\bar{\tb{C}}$.  The invariants
of $\bar{\tb{C}}$ and $\tb{C}$ are related in the following way,
\begin{equation}
    \bar{I}_1=\exp\left(s\right)I_1;\;\;\;\bar{I}_2=\exp\left(2s\right)I_2;\;\;\;\bar{I}_3=\exp\left(3s\right)I_3,
\end{equation} 
Here, $I_i$ and $\bar{I}_i$ are the principal invariants of $\tb{C}$ and
$\bar{\tb{C}}$ respectively.  In the literature, the Green-Lagrangian strain
tensor is sometimes defined as half the difference between the deformed and the
reference metrics (see \cite{green1992}, \cite{efrati2009}).  Although this
definition might make sense in a small deformation setting, it is problematic
since the addition of metric tensors is not well defined. To be more precise,
the reference metric operates only on tangent vectors of the reference
configuration, while the deformed metric operates on tangent vectors of the
deformed configuration. In some cases, by the 'deformed metric', the pulled-back
metric is implied.  The distinction between the deformed metric and the pulled
back metric is quite important in a geometrically motivated continuum theory,
since the metric tensor may be construed as a separate dynamical field
describing the intrinsic geometry of the body.  

The Jacobian of deformation in the presence of temperature change is denoted by
$\bar{J}$ and is given by,
\begin{equation}
    \bar{J}=\exp\left(\frac{3s}{2}\right)\sqrt{\frac{g}{G}}\det{\tb{F}}
\end{equation}
where, $g$ and $G$ denote the determinants of the reference and deformed metric
tensors. The non-metricity tensors of the reference and deformed configurations
are denoted by $\tb{H}$ and $\tb{Q}$ respectively. These tensors are defined by,
\begin{equation}
    \tb{H}=\xi \otimes \tb{G};\;\;\; \tb{Q}=\phi \otimes \tb{g}
\end{equation}
In this work, the non-metricity tensor of the deformed configuration completely
characterizes the residual stresses developed in the body due to inelastic
effects. The initial configuration may also be trivially viewed as a deformed
configuration with identity map as deformation. With this understanding, the
tensor $\tb{H}$ has significance well beyond what the metric tensor of the
initial configuration offers. Specifically, as a conjugate to the kinematic
variable $\xi$, one may talk about residual stresses using just one
configuration. This may be contrasted with elastic stresses (stresses due to
deformation) whose description necessitates both the reference and deformed
configurations.

The pull-back of the deformed configurations Weyl one-form, denoted as
$\phi'\in T^*\mathcal{B} $, is given by,
\begin{equation}
    \phi'=\bs{\varphi}^*\phi
\end{equation}
Similarly, the pulled-back non-metricity tensor $\tb{Q}'$ is defined by
$\tb{Q}':=\bs{\varphi}^*(\tb{Q})$, which leads to the following expression,
\begin{equation}
    \tb{Q}'=\phi'\otimes \bar{\tb{C}}
\end{equation}
This expression is obtained by applying the pull-back map to the Weyl
one-form and the metric tensor.  Similarly, the pull-back of the distant
curvature is denoted by $\tb{K}':=\bs{\varphi}^*(\tb{K})$. Since the pull-back
map commutes with exterior derivative, we can write $\tb{K}'$ as,
\begin{align}
    \nonumber
    \tb{K}' &=\bs{\varphi}^*d \phi\\
    \nonumber
    &=d(\bs{\varphi}^* \phi)\\
    &=d \phi'
\end{align}
The spatial rate of deformation tensor $\tb{d}$, characterizes the relative
velocity of the material in a small neighbourhood around a point in the deformed
configuration and is given by,
\begin{equation}
    \tb{d}=\frac{1}{2}\mathcal{L}_{\tb{v}}\tb{g}
    \label{eq:rateofDeformation}
\end{equation}
where, $\mathcal{L}_{\tb{v}}(.)$ denotes the Lie derivative. When the geometry
of the material body is Riemannian, the formula above reduces to
$d_{ij}=\frac{1}{2}(v_{i;j}+v_{j;i})$. An index after a semicolon denotes the
covariant derivative (given by the appropriate connection). The Lie derivative of
the metric tensor is,
\begin{equation}
    \mathcal{L}_{v^c \partial_c} (g_{ab})=v^c\partial_c
    g_{ab}+g_{cb}\partial_av^c +g_{ca}\partial_bv^c
\end{equation}
By adding and subtracting $\gamma_{da}^c v^d$ and $\gamma_{db}^cv^d$ in the
last equation and using the definition of covariant derivative, we arrive at,
\begin{equation}
    \mathcal{L}_{v^c \partial_c} (g_{ab})=v^c \nabla_c g_{ab} +g_{cb}\nabla_av^c
    +g_{ca}\nabla_b v^c
\end{equation}
Now, using the properties of Weyl connection, the equation above may be rewritten
as,
\begin{align}
    \mathcal{L}_{v^c \partial_c} (g_{ab})&=\phi_c v^c g_{ab}+g_{cb} \nabla_a v^c
    + g_{ca} \nabla_b v^c\\
    &=\phi_cv^cg_{ab}+ \nabla_a v_b+\nabla_b v_a-(\phi_a v_b+ \phi_b v_a)
    \label{eq:lieDerivativeMetric}
\end{align}
where, $v_a=g_{ab}v^b$ and the relation $\nabla_a v_c= \nabla_a g_{cb} v^c+
g_{cb}\nabla_av^c$ have been used.
Using Eq. \ref{eq:lieDerivativeMetric}, the rate of deformation tensor may now
be computed. Substituting Eq. \ref{eq:lieDerivativeMetric} in Eq.
\ref{eq:rateofDeformation}, we arrive at rate of deformation tensor of a body
whose configurational geometry is Weylian.
\begin{equation}
    d_{ij}=\frac{1}{2}\left(v_{i;j}+v_{j;i}-(\phi_i v_j+ v_i\phi_j)+\phi_kv^k
    g_{ij}\right)
\end{equation}
It should be noted that the rate of deformation tensor for a body with Weylian
geometry  has three additional terms compared with the Riemannian
case. These additional terms are a consequence of the incompatibility of the
connection with the metric.

\subsection{Connections of reference and deformed configurations}
The connection of the reference configuration is assumed to be time independent,
while that of the deformed configuration evolves with time. The connection
coefficients of reference and deformed configurations are denoted by
$\Gamma_{AB}^{C}$ and $\gamma_{ab}^c$ respectively. If the reference
configuration has a non-trivial connection (different from the flat
Levi-Civita), then the body has a distribution of defects, which may require
residual stresses for mechanical equilibrium to be maintained. The connection of
the deformed configuration is determined by the evolution of the Weyl one-form,
which in turn is coupled to the evolving temperature field.  This evolution
models the generation, motion and extinction of metrical defects due to
externally applied thermo-mechanical stimuli.

\section{Kinetics}
The balance of linear momentum in the reference configuration is given by,
\begin{equation}
    \nabla.\tb{P}+\tb{B}=\rho_{0}\frac{d\tb{V}}{dt}
    \label{eq:mechEquilibrium}
\end{equation}
Here, $\tb{P}$ is the first Piola stress and $\nabla(.)$ is the divergence
operator determined by the Weyl connections of the reference and deformed
configurations. The Piola stress $\tb{P}$ is related to the Cauchy stress
$\bs{\sigma}$ through the Piola transform (which is applied to the second index
of the Cauchy stress). We define the symmetric Piola (second Piola) stress as
the pull-back of the first index of the first Piola stress. The relation between
the Cauchy stress and second Piola stress is thus given by,
\begin{equation}
    \tb{S}=J \bs{\varphi}^*(\bs{\sigma})
\end{equation}

\subsection{Evolution of Weyl one-form}
\label{sec:evolWeylOneForm}
The equilibrium condition for the Weyl one-form is obtained as a critical point
of the free-energy with respect to the one-form $\phi$. In the reference
configuration, this condition may be written as,
\begin{equation}
    \delta_{\phi}\hat{\psi}_{0}=0;\;\;\;
    \delta_{\phi}\hat{\psi}_{0}:= \delta_{\phi}\int_{\mathcal{B}} \psi_{0}dV
    \label{eq:minimumOneform}
\end{equation}
where, $\hat{\psi}_{0}$ is the free energy of the body and $\delta_{\phi}(.)$
denotes the G\^ateaux or the variational derivative.  Assuming that the free
energy depends on $\tb{F}$, $\tb{Q}'$, $\tb{K}'$ and $T$, its variation may be
evaluated as,
\begin{equation}
    \delta_{\phi} \psi_{0}=\frac{\partial \psi_{0}}{\partial
    \tb{F}}:\delta_\phi \tb{F}+\frac{\partial \psi_{0}}{\partial
    \tb{Q}'}\dot{:}\delta_\phi \tb{Q}'+  \frac{\partial \psi_{0}}{\partial
    \tb{K}'}:\delta_\phi \tb{K}'
\end{equation}
$\delta_\phi(\tb{A})$ denotes the directional derivative of the tensor $\tb{A}$
with respect to $\phi$. Since the deformation gradient   does not depend on
$\phi$, we have $\delta_{\phi} \tb{F}=\tb{0}$. We also have $\delta_\phi
\tb{K}=d\delta\phi$ and hence $\delta_\phi \tb{K}'=d(\delta\phi)'$; here
$(\delta \phi)':=\bs{\varphi}^*(\delta \phi)$. Similarly,
$\delta_{\phi}\tb{Q}'=(\delta \phi)' \otimes \bar{\tb{C}}$. We also introduce
the following definitions for a simpler exposition,
\begin{equation}
    \tb{M}:=\frac{\partial \psi_{0}}{\partial \tb{Q}'};\;\;\;
    \tb{N}:=\frac{\partial \psi_{0}}{\partial \tb{K}'}
    \label{eq:defMandN}
\end{equation}
Using these results and definitions, the variation of the free energy may be
written as,
\begin{equation}
    \int_{\mathcal{B}} \left(M^{IJK}C_{JK} (\delta
    \phi)_{I}'+N^{IJ}\partial_{[I}(\delta
    \phi)'_{J]} \right)dV=0
\end{equation}
Applying the relation between $(\delta \phi)'$ and $\delta \phi$ and using the
divergence theorem on the second term, we arrive at,
\begin{equation}
    \nabla_J N^{IJ} -M^{JKI}\bar{C}_{JK} =0
    \label{eq:defectEquilibrium}
\end{equation}

\subsection{Doyle-Ericksen formula}
In continuum mechanics, the Doyle-Ericksen formula \cite{doyle1956, mfe} is an
important result, which relates stress (Cauchy stress) with the metric of the
deformed configuration. Through the metric, this formula is thus able to relate
the intrinsic geometry of the deformed configuration to the stresses developed.
The Doyle-Ericksen formula is given by,
\begin{equation}
    \sigma^{ij}=2\rho\frac{\partial \psi}{\partial g_{ij}}
    \label{eq:DoyleErichsenFormula}
\end{equation}
$\rho$ and $\psi$ are the mass density and free energy density of the deformed
configuration. An interesting aspect of the formula is that it emphasizes length
as a fundamental quantity. If the last statement is understood abstractly, then
there is no reason to restrict length just to its Euclidean notion. To obtain a
relationship between the Cauchy stress and the scaled metric $\bar{\tb{g}}$, we
replace $\tb{g}$ in Eq. \ref{eq:DoyleErichsenFormula} by $\bar{\tb{g}}$, so the
desired relation is given by,
 \begin{equation}
     \sigma^{ij}=2\rho\frac{\partial \psi}{\partial \bar{g}_{ij}}
    \label{eq:DoyleErichsenScaled}
 \end{equation}
In the reference configuration, the equation above provides a relation between
the second Piola stress and the right Cauchy-Green deformation tensor, 
\begin{equation}
    S^{IJ}=2\rho_{0}\frac{\partial \psi_{0}}{\partial \bar{C}_{IJ}}
\end{equation}
Note that this equation is related to Eq.  \ref{eq:DoyleErichsenScaled}.  To
prove the equivalence, one starts by noticing that,
\begin{equation}
    \varphi_*\frac{\partial \psi_{0}}{\partial
    \bar{C}_{IJ}}=\frac{\partial \psi}{\partial \bar{g}_{ij}}
    \label{eq:intermediateRelation}
\end{equation}
Moreover, the second Piola stress is related to the Cauchy stress though the
push-forward of its first index and an inverse Piola transform applied to the
second index. Formally, these two operations may be written together as,
\begin{equation}
    \sigma^{ij}=\frac{1}{J}\varphi_*(S^{IJ})
    \label{eq:relationSecondPiolaCauchy}
\end{equation}
Eqs. \ref{eq:intermediateRelation} and \ref{eq:relationSecondPiolaCauchy}
together lead us to the required equivalence. In the present context, the notion
of stress is understood in the sense of Eq.  \ref{eq:DoyleErichsenFormula}. The
positive scalar factor introduced in the metric by the Weyl transformation does
not create a new notion of stress, unlike what was conceived of by Gurtin.
Gurtin in \cite{gurtin1996} postulated the existence of new stresses called
micro-stresses and a new balance rule called the micro-force balance.  These
stresses were shown as energetically conjugate to the internal variables
introduced to describe the inelastic deformation. Moreover, the divergence-type
balance law for the internal variables was restrictive and a physical meaning
for these micro-stresses remained elusive. Identifying thermal strains as a
geometric object permits us to reinterpret the classical geometric results
seamlessly. This is often impossible with internal variable theories, which
require additional postulates for closure.

\subsection{Comparision with Duhamel-Neuman relation}
In linear thermo-elasticity, the Duhamel-Neuman hypothesis is commonly adopted
to model the mechanical response of solids in the presence of temperature
change. It postulates the existence of a new tensor-valued internal variable
called the thermal strain whose evolution is directly proportional to the
temperature change in the body.  The total stain $\bs{\epsilon}_{\text{Tot}}$,
defined as the symmetric part of the displacement gradient, is recovered by an
additive splitting,
\begin{equation}
    \bs{\epsilon}_{\text{Tot}}=\bs{\epsilon}_{\text{Mec}}+\bs{\epsilon}_{\text{The}}.
\end{equation}
The mechanical and thermal parts of the strain are denoted by
$\bs{\epsilon}_{\text{Mec}}$ and $\bs{\epsilon}_{\text{The}}$ respectively. For
a body admitting isotropic thermal expansion, $\bs{\epsilon}_{\text{The}}$
relates to the change in temperature $(T-T_0)$ in the following way,
\begin{equation}
    \bs{\epsilon}_{\text{The}}=\alpha (T-T_0)\tb{I}
\end{equation}
Several extensions of this additive splitting to finite deformation
thermo-elasticity exist in the literature. Splitting of the deformation gradient
in thermal and elastic parts comes in two forms: $\tb{F}=\tb{F}_e\tb{F}_T$ and
$\tb{F}=\tb{F}_T\tb{F}_e$. Implicit in these settings is the notion of an
intermediate configuration to describe the incompatibility created by thermal
deformation. Yet another extension of the Duhamel-Neuman hypothesis to finite
deformation is through an additive decomposition of the rate of deformation
tensor into mechanical and thermal components $\tb{d}=\tb{d}_e+\tb{d}_T$ (for
details see \cite{mfe}), where,
\begin{equation}
    \tb{d}=\frac{1}{2}\rho_0\left[ \left(\frac{\partial^2 \chi }{\partial \bs{\tau}^2}
    :\mathcal{L}_{\tb{v}} \bs{\tau}\right)+\frac{\partial^2 \chi}{\partial
    \bs{\tau} \partial T}\dot{T}\right].
    \label{eq:finiteThermoElasSplit}
\end{equation}
and $\tb{d}_e$ and $\tb{d}_T$ are given as,
\begin{equation}
    \tb{d}_e= \frac{\rho_0}{2}\left(\frac{\partial^2 \chi }{\partial \bs{\tau}^2}
    :\mathcal{L}_{\tb{v}} \bs{\tau}\right);\;\;\;
    \tb{d}_T= \frac{\rho_0}{2} \frac{\partial^2 \chi}{\partial
    \bs{\tau} \partial T}\dot{T}
    %+\rho_0\frac{\partial^2 \chi }{\partial \bs{\tau} \partial{T} }\right)
\end{equation}
Here, $\chi$ denotes the complementary free energy and $\bs{\tau}$ the Kirchhoff
stress. The thermo-elastic splitting given in Eq. \ref{eq:finiteThermoElasSplit}
is verifiable using Legendre transform; hence it is not a hypothesis. It only
requires the existence of a temperature dependent free energy.  In any case, the
splitting of the rate of deformation tensor cannot account for the
incompatibility due to thermal strain since the geometry of the body remains
essentially Euclidean. 

With our present formulation, we are able to avoid the notion
of an intermediate configuration whilst incorporating the incompatibility
due to thermal strains in a consistent manner. The rate of deformation tensor
$\bar{\tb{d}}=\frac{1}{2}\mathcal{L}_{\tb{v}} \bar{\tb{g}} $ may be written as,
\begin{equation}
    \bar{\tb{d}}= e^s\tb{d}  + \frac{1}{2}\tb{v}[s]\bar{\tb{g}}
\end{equation}
The equation above follows from the  properties of the Lie derivative, $\tb{v}[.]$
denotes the action of a vector field on a function,
$\tb{d}:=\mathcal{L}_{\tb{v}}\tb{g}$ and
$\bar{\tb{d}}:=\mathcal{L}_{\tb{v}}\bar{\tb{g}}$. 
The current geometric methodology has the added advantage that the rate of
deformation tensor may be determined kinematically, without the use of inverse
elastic relations. Neither complementary free energy nor Legendre transform
is used to arrive at the equation. However, the decomposition given in
Eq.  \ref{eq:finiteThermoElasSplit} may also be adopted in the present setting.

\section{Thermodynamics}
We use a local equilibrium thermodynamics framework to determine the
restrictions imposed on the constitutive functions by the laws of
thermodynamics. In line with the standard hypothesis, we postulate the existence
of state variables called the specific internal energy density and specific
entropy, denoted by $U$ and $\eta$ respectively. $U$ is assumed to depend on the
right Cauchy-Green deformation tensor, non-metricity tensor, distant curvature
and specific entropy. $U$ may also depend on co-ordinates of the reference
configuration; however we choose to work with an internal energy which is
homogeneous, i.e. the explicit spatial dependence is ignored. The balance of
energy, which is the statement of the first law of thermodynamics, may be
written as,
\begin{equation}
    \frac{d}{dt}\int_{\mathcal{B}}\rho_{0}\left(U+\frac{1}{2}\tb{G}(\tb{V},\tb{V})\right)dV=
    \int_{\mathcal{B}}\langle \tb{B},\tb{V}\rangle dV+\int_{\partial
    \mathcal{B}}\langle \tb{t},\tb{V} \rangle dA-\int_{\partial
    \mathcal{B}}\langle \tb{q},\tb{n} \rangle dA
    \label{eq:GlobalBalanceEnergy}
\end{equation}
In the above equation, $\langle.,.\rangle$ denotes the natural pairing between
cotangent and tangent vectors. The heat flux vector and the unit vector normal
to the boundary $\partial \mathcal{B}$ are denoted by $\tb{q}$ and $\tb{n}$
respectively.  Kanso \textit{et} \textit{al.} \cite{kanso2007} have interpreted
the stress tensor as a bundle valued two-form. Such a description has the
advantage of having the integrals in the balance of energy consistent with the
integration of differential forms.  However, introducing such notions of stress
do not affect the resulting equations; hence we do not dwell on such technical
details here.  The temperature gradient of the deformed configuration is given
by $\bar{g}^{ij}\partial_jT$. To determine the heat flux vector of the reference
configuration, one may adopt one of the following two routes. In the first, one
exploits the constitutive relation between temperature gradient and heat flux
in the deformed configuration and use the Piola transform to bring the heat flux
vector to the reference configuration. Alternatively, one may use the Piola
transform on the temperature gradient of the deformed configuration and use the
referential version of the constitutive rule connecting heat flux and
temperature gradient. The two routes are equivalent if the constitutive rule is
transformed carefully.  

The equilibrium equation for the one-form given in Eq.
\ref{eq:defectEquilibrium} does not account for time dependent relaxation
experienced by defects.  We model time dependent relaxation by adjoining the
equilibrium equation for the Weyl one-form with an additional viscous term. Such
a procedure to obtain the evolution rule for states that relax to equilibrium
has been employed in arriving at the Cahn-Hillard and Ginsberg-Landau equations.
Geometric evolutions like the Ricci flow and mean curvature flow equations also
have a similar variational structure. The equilibrium equation for the one-form
with relaxation may be written as,
\begin{equation}
    \mathcal{L}_{\tb{V}}(\phi')=\frac{1}{m}\delta_{\phi'} \hat{\psi}_{0}
\end{equation}
$\mathcal{L}_{\tb{V}}(.)$ denotes the Lie derivative with respect to the
velocity field, $m$ is a constant describing the time dependent relaxation
experienced by defects. In addition to the balance of energy for the whole body,
we also postulate that a local form of energy balance holds (the fields used in
the integral statement of Eq. \ref{eq:GlobalBalanceEnergy} are assumed
sufficiently smooth so that localization theorem can be applied), which may be
written as,
\begin{equation}
    \rho_{0}\dot{U}=\frac{1}{2}S^{IJ}\dot{\bar{C}}_{IJ}+N^{IJ}\dot{\phi}'_{[I,J]}+M^{IJK}\bar{C}_{JK}\dot{\phi}_I'-\partial_I q^I +\rho_0 R-m \dot{\phi}^{'I} \dot{\phi}'_I 
    \label{eq:localEnergyBalance}
\end{equation}
where, $R$ denotes the heat source and $(.)_{[I,J]}$ the anti-symmetrization
with respect to the indices $I$ and $J$. The last equation is obtained from Eq.
\ref{eq:GlobalBalanceEnergy} through the use of the balance of linear momentum,
divergence theorem and localization theorem. We impose the second law of
thermodynamics through the Clausius-Duhem inequality; its local form is given
by,
\begin{equation}
    \rho_{0}\dot{\eta} \geq
    \rho_{0}\frac{R}{T}-\partial_I\left(\frac{q^I}{T}\right)
    \label{eq:localClacusiusDuhem}
\end{equation}
Applying the Legendre transform $\psi_{0}=U-T\eta$ with respect to the
conjugate variables $\eta$ and $T$, we arrive at,
\begin{equation}
    \dot{\eta}=\frac{1}{T}(\dot{U}-\dot{\psi}_0-\dot{\theta}\eta);\;\;\;
    %\label{eq:entropyrate}
\eta=-\frac{\partial \psi_{0}}{\partial T}.
\label{eq:conjTempEntropy}
\end{equation}
The reference free energy density is assumed to be a function of the right
Cauchy-Green deformation tensor, non-metricity tensor, distant curvature tensor
and temperature. The rate of Helmholtz free energy density is calculated as,
\begin{equation}
    \dot{\psi}_0=\frac{\partial \psi_0}{\partial
    \bar{C}_{IJ}}\dot{\bar{C}}_{IJ}+\frac{\partial \psi_0}{\partial
    Q'_{IJK}}\bar{C}_{JK} \dot{\phi}'_I+\frac{\partial \psi_0}{\partial
    Q'_{IJK}}\phi_I'\dot{\bar{C}}_{JK}+\frac{\partial \psi_0}{\partial
    K'_{IJ}}\dot{\phi}'_{[I,J]}+\frac{\partial \psi_0}{\partial T}
    \label{eq:rateFreeEnergy}
\end{equation}
In arriving at the equation, use is made of the relation
$\dot{Q}_{IJK}=\dot{\phi}'_I\bar{C}_{KL}+\phi_I\dot{\bar{C}}_{KL}$. 
Using the relationship between entropy, internal energy and free-energy density
in Eq. \ref{eq:localClacusiusDuhem}, we have the following form of second law,
\begin{equation}
    \rho_0(\dot{U}-\dot{\psi}_0-\dot{T}\eta)+\partial_Iq^I-\frac{q^I}{T}\partial_I
    T-\rho_0R \geq 0 
    \label{eq:finalCaluciusDuhem}
\end{equation}
Now, substituting the rate of free energy and internal energy in Eq.
\ref{eq:finalCaluciusDuhem} leads to,
\begin{eqnarray}
    \nonumber
    \left(\frac{1}{2}S^{IJ}-\rho_0\left(\frac{\partial\psi_0}{\partial\bar{C}_{IJ}}+\frac{\partial
    \psi_0}{\partial
    Q'_{KIJ}}\phi'_K
    \right)\right)\dot{\bar{C}}_{IJ}+\left(M^{IJK}-\rho_0\frac{\partial \psi_0}{\partial
    Q_{IJK}'}\bar{C}_{JK} \right)\dot{\phi}'_I\\
    +\left(N^{IJ}-\rho_0\frac{\partial \psi_0}{\partial K'_{IJ}}\right)
    \dot{\phi}'_{[I,J]}-\left(\frac{\partial
    \psi_0}{\partial T}+\eta \right)\dot{T} -m\dot{\phi}^{'I}\dot{\phi}'_I-\frac{q^I}{T}\partial_IT \geq 0
    \label{eq:colemanNoll}
\end{eqnarray}
%\begin{equation}
%    \dot{\psi}=\frac{\partial \psi_{\text{Ref}}}{\partial
%    \bar{\tb{C}}}:\dot{\bar{\tb{C}}}
%+\frac{\partial \psi_{\text{Ref}}}{\partial T} \dot{T}
%\label{eq:psiDot}
%\end{equation}
%in the first law, leads to the following equation,
%\begin{equation}
%\rho_{\text{Ref}}T\dot{\eta}=-\left(\rho_{\text{Ref}}\frac{\partial
%    \psi_{Ref}}{\partial \bar{\tb{C}}}-\tb{S}\right):\dot{\bar{\tb{C}}}
%-\left(\rho_{\text{Ref}}\frac{\partial \psi}{\partial T}+s\right)\dot{T}+\nabla.\tb{q}-\rho R
%\label{eq:FirstLawAfterLegenderTransform}
%\end{equation}
Applying a Coleman-Noll type procedure to the last equation gives us the
requisite constitutive relations. Thus the second Piola stress is given by,
\begin{equation}
    S^{IJ}=2\rho_0\left(\frac{\partial \psi_0}{\partial \bar{C}_{IJ}}+\frac{\partial \psi_0}{\partial Q'_{KIJ}}\phi'_K \right)
    %\tb{S}=2\rho_{\text{Ref}}\frac{\partial \psi_{\text{Ref}}}{\partial
    %\bar{\tb{C}}}
    \label{eq:sPkConstitution}
\end{equation}
Similarly, we also have,
\begin{equation}
    M^{IJK}=\frac{\partial \psi_0}{\partial Q'_{IJK}};\;\;\;
    N^{IJ}=\frac{\partial \psi_0}{\partial K'_{IJ}}
\end{equation}
The last two relations were established through a variational argument in
the previous Section \ref{sec:evolWeylOneForm}. Using these constitutive relations in the dissipation inequality leads to,
\begin{equation}
    -\left(m\dot{\phi}^{'I}\dot{\phi}'_I+\frac{q^I}{T}\partial_IT\right) \geq 0
\end{equation}
From this, it follows that $m<0$ and $q^I=-L^{IJ} \partial_J T$,
where $\tb{L}$ is the thermal conductivity which is positive definite.

\subsection{Temperature evolution}
The temperature evolution equation is obtained by substituting the constitutions for
second Piola stress and entropy into the local form of energy balance.
Using Eq. \ref{eq:conjTempEntropy}, the rate of entropy production $\dot{\eta}$
is given by,
\begin{equation}
    \dot{\eta}=-\left(\frac{\partial^2 \psi_{0}}{\partial
    T^2}\dot{T}+\frac{\partial^2 \psi_{0}}{\partial \bar{\tb{C}}
    \partial T}:\dot{\bar{\tb{C}}}+\frac{\partial^2 \psi_0}{\partial
    \bar{\tb{Q}}'
    \partial T}\dot{:}\dot{\bar{\tb{Q}}}'+\frac{\partial^2 \psi_0 }{\partial
    \bar{\tb{K}}' \partial
    T}:\dot{\bar{\tb{K}}}' \right)
    \label{eq:entropyRate}
\end{equation}
substituting Eqs. \ref{eq:entropyRate}, \ref{eq:conjTempEntropy} and
\ref{eq:sPkConstitution} into the energy balance leads to the following
evolution equation for temperature.
\begin{equation}
    -\rho_{0} T\left(\frac{\partial^2 \psi_{0}}{\partial
    T^2}\dot{T}+\frac{\partial^2 \psi_{0}}{\partial \bar{\tb{C}}
    \partial T}:\dot{\bar{\tb{C}}}+\frac{\partial^2 \psi_0}{\partial
    \bar{\tb{Q}}'
    \partial T}\dot{:}\dot{\bar{\tb{Q}}}'+\frac{\partial^2 \psi_0 }{\partial
    \bar{\tb{K}}' \partial
    T}:\dot{\bar{\tb{K}}}' \right)
=
    -\nabla.\tb{q}+\rho_{0}R
\end{equation}

\subsection{Residual stresses due to defects}
An accurate prediction of residual stresses requires complete information on
the distribution of defects. As assumed, the Weyl
one-form of the deformed configuration encodes the metrical defects present in
the body. This one-form may or may not evolve, depending on the
thermo-mechanical processes the body is subjected to. Assume the initial
configuration $\mathcal{B}$ as unloaded and without deformation, but with
defects. The no-deformation assumption implies that $\tb{F}=\tb{I}$, whereupon
it follows that Cauchy, first and second Piola stresses are
indistinguishable or, in other words, the Piola transform and pull-back operation
are trivial (identities).  
\begin{equation}
    \tb{P}=\tb{S};\;\;\;\tb{S}=\bs{\sigma}
\end{equation}
The condition of no external traction or displacement on $\mathcal{B}$ implies that
the residual stresses have to be self equilibrating.  
\begin{equation}
    \sigma^{IJ}_{\;\;\;\;;J}=0
\end{equation}
In addition, if  we assume that residual stresses are obtainable from a
stored energy function, then the stress tensor may be written as,
\begin{equation}
    \sigma^{IJ}=2\rho_{0}\frac{\partial \psi^R}{\partial G_{IJ}}
    \label{eq:freeEneResStrNoDef}
\end{equation}
where, $\psi^R$ is the free energy due to defects.  The last
equation is the Doyle-Ericksen formula discussed in the previous section. The
metric tensor appearing in the formula is one on
$\mathcal{B}$.  The assumption that residual stresses are characterizable
using a free energy, is grounded in the physical reality that a 
residually stressed body, on being cut, would deform. This implies that the energy stored in the body due to the
presence of defects may be converted to strain energy through a deformation
process. This property is often used in an experimental characterization of
residual stresses through destructive testing. It should be noted that the
free energy for the residual stress field need not be the same as the stored
energy used to determine a deformation process. Distributions of the Weyl
one-form and the stored energy $\psi^R$ together determine the residual stress
distribution on $\mathcal{B}$. 

When $\mathcal{B}$ is subjected to deformation, the defects present in the
body evolve, which along with the deformation-induced stresses
equilibrates the externally applied loads. Thus we
postulate that the free energy of the body at a configuration has two components: one
due to deformation and temperature, and the other due to defects. Using a
referential description, the component of free energy due to defects is written
as a function of the pulled back non-metricity tensor and the pulled back
distant curvature; we denote this component by $\psi_0^R$.  The deformation and
temperature dependent component of free energy is denoted by $\psi_0^{\varphi}$;
it is assumed to be a function of the deformation gradient, metric tensor of the
deformed configuration and temperature.
\begin{equation}
    \psi_{0}=\psi_{0}^{\varphi}(\tb{F},\tb{g},T)+
    \psi^R_0(\tb{Q}',\tb{K}')
    \label{eq:freeEnergyWithResidualStress}
\end{equation}
Assuming the Doyle-Ericksen formula to hold in the deformed configuration, the
second Piola stress is now given by,
\begin{equation}
    \tb{S}=\frac{\partial \psi^\varphi_{0}}{\partial \bar{\tb{C}}}+\frac{\partial
    \psi^R_{0}}{\partial \tb{Q}'} \frac{\partial \tb{Q}'}{\partial
    \bar{\tb{C}}}
\end{equation}
The first term on the right hand side is the second Piola stress caused by 
deformation, while the second term is due to the defects.  Note that, when the Weyl one-form vanishes, 
so does the defect component of the free
energy leading to the usual expression for the second Piola stress.
Similarly, when deformation vanishes, $\tb{F}=\tb{I}$ and the right Cauchy-Green
deformation tensor reduces to the metric of the deformed configuration leading
to Eq. \ref{eq:freeEneResStrNoDef}.

\subsection{A specific constitutive rule}
\label{sec:specificConstitution}
Having dwelt on the thermodynamic framework and the form of the constitutive
rules in the previous subsections, we now make a specific choice for the constitutive
relations. We assume the deformation part of the free energy to be of the
compressible Neo-Hookian type. Indeed, any hyperelastic
free energy function could be used in its place. The stored energy for a
compressible neo-Hookian material is given by,
\begin{equation}
    \psi_0^{\varphi}=\frac{\mu}{2}(\bar{I}_1-3)-\mu \log(\bar{J})+\frac{\lambda}{2}\log(\bar{J})^2
    \label{eq:storedEnergy}
\end{equation}
Note that the principal invariants used in the stored energy density are Weyl
transformed. For the defect part of the free energy, we assume the following
form,
\begin{equation}
    \psi_0^R=\frac{1}{2}\left(k_1\bar{Q}'_{IJK}G^{IL}G^{JM}G^{KN}\bar{Q}'_{LMN}
    +k_2G^{KM}G^{LN}K'_{KL}K'_{MN}\right)
\end{equation}
In the last equation, $k_1$ and $k_2$ are material constants
characterizing the defect-induced free energy. The 
defect free energy may be further simplified as,
\begin{equation}
    \psi_0^R=\frac{1}{2}\left(k_1 \bar{C}_{IJ}\bar{C}^{IJ}\phi'_K
    {\phi'}^{K}+k_2{K'}^{MN}K'_{MN}\right)
\end{equation}
In the above expression, the definitions $\bar{C}^{IJ}=\bar{C}_{KL}G^{IK}G^{JL}$ and
${K'}^{IJ}=G^{IK}G ^{JL}{K'}_{KL}$ are used. The second Piola stress generated
from such a free energy is given by,
\begin{equation}
    S^{IJ}=\frac{\mu}{2}G^{IJ}+\frac{1}{2}(-\mu+ \lambda
    \log(\bar{J}))(\bar{C}^{-1})^{IJ}+k_1\bar{C}^{IJ}{\phi'}^K \phi'_K
\end{equation}
Using Eq. \ref{eq:defMandN}, tensors $\tb{M}$ and $\tb{N}$ for the assumed
free energy function are given by,
\begin{equation}
    M^{IJK}=k_1\bar{C}^{IJ} {\phi'}^K;\;\;\;
    N^{IJ}=k_2{K'}^{IJ}
\end{equation}

\section{Stresses due to a point defect}
\label{sec:pointDefect}
We present the calculation of self stresses due to a point defect using the
geometric theory considered in the preceding sections. Presently, we ignore
defect evolution and focus only on the mechanical equilibrium. This description
exploits the fact that the equilibrium configuration of a body with a nontrivial
Weyl connection is different from that with a flat Levi-Civita connection. For a
body with a hyperelastic stored energy and flat Levi-Civita connection, the
identity deformation is always a critical point under zero traction and zero
boundary displacement. In other words, the reference configuration (identity
deformation) is a natural state. In the present approach, the identity
deformation is stress-free but not the only critical point of the stored energy
function: we conjecture that this critical point is unstable when the body is
defective. In the presence of point defects, the connection is modified locally
and the Weyl one-form encodes this information. 

In the linear elasticity setting, point defects are analogous to a spherical
inclusion forced into a spherical cavity of slightly different diameter. The
difference in the volume of the cavity and inclusion results in stresses in the
inclusion and the matrix surrounding it. The infinitesimal version of the
inclusion problem introduces the notion of a dipole tensor to characterize the
pre-stress caused by the point defect. This description may be referred to as
Eshelby's method of eigenstrain in the infinitesimal setting.  For a recent
review on modelling point defects based on the elasticity theory, see
\cite{clouet2018}.  The accuracy of linear elastic predictions for point defects
depends on how closely the dipole tensor represents the point defect and the
dependence of the dipole tensor on the energy functional. Different approaches
have been suggested for obtaining the dipole tensor from molecular dynamic
simulations of point defects. These methods use the atomistic stress,
displacement or the Kansaki force to deduce the elastic dipole tensor. A main
disadvantage of the linear elastic approach to point defects is that it cannot
be extended to include nonlinear interactions. 

\begin{figure}
    \centering
    \includegraphics[width=0.5\textwidth]{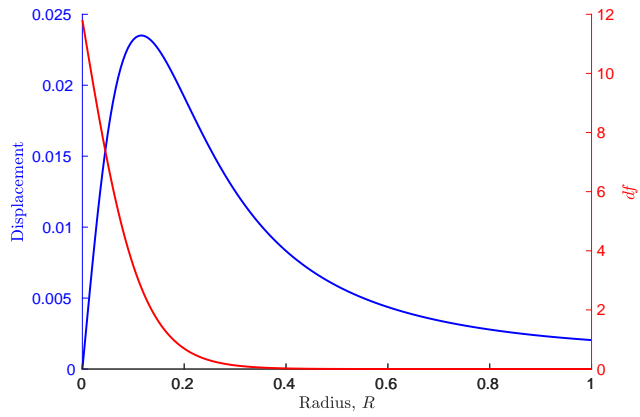}
    \caption{Radial displacement at equilibrium for the assumed one-form $df$}
    \label{fig:oneForm_Displacement}
\end{figure}

\begin{figure}
    \centering  
    \begin{subfigure}[b]{.90\textwidth}
        \centering
        \includegraphics[width=0.5\textwidth]{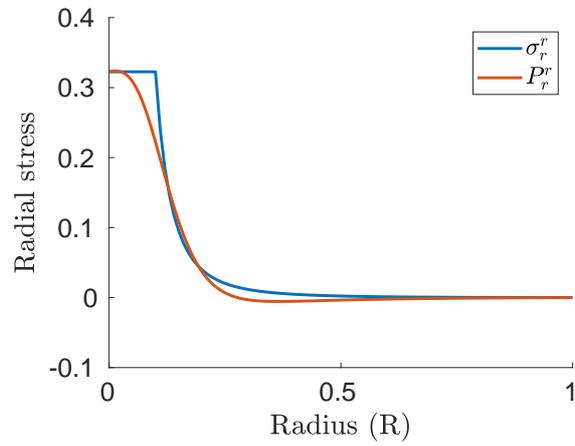}
        \caption{Radial stress}
    \end{subfigure}
    \begin{subfigure}[b]{.90\textwidth}
        \centering
        \includegraphics[width=0.5\textwidth]{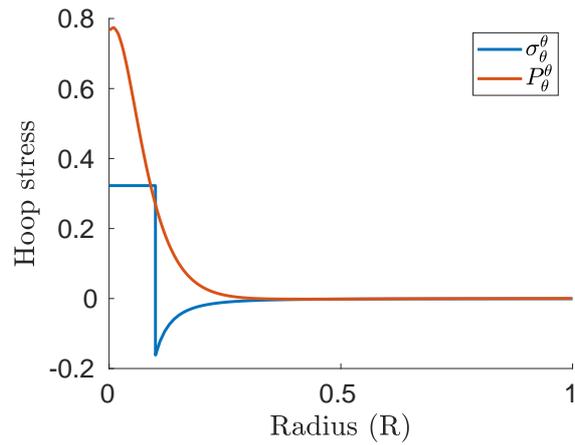}
        \caption{Hoop stress}
    \end{subfigure}
    \caption{Comparisons of radial and hoop stresses computed by the present
    method and linear elasticity. The linear elastic solution was obtained
    from \cite{teodosiu2013} }
    \label{fig:radialHoopstress}
\end{figure}

We now consider a spherical body with radius one and a point defect placed at
the origin. As opposed to linear elasticity, the point defect is modelled by a
one-form which is localized near the origin and reaches zero asymptotically away
from the origin. This description is similar to a shrink-fit problem with a
diffused shrink-fit surface. A spherical coordinate system is used for the
reference and deformed configurations of the body; these coordinates are denoted
by $(R, \Theta,\Xi)$ and $(r, \theta, \xi)$ respectively.  We adopt the stored
energy function given in Eq.  \ref{eq:storedEnergy} and set the coefficient
$k_1$ to zero. The Weyl one-form is assumed to be integrable. The metric tensor
in the spherical coordinate system is given by, 
\begin{equation}
    G_{IJ}=\begin{bmatrix}
        1 & 0 & 0\\
        0 & R^2\sin^2 \Xi & 0\\ 
        0 & 0 & R^2\\ 
    \end{bmatrix};\;\;\; 
    g_{ij}=\begin{bmatrix}
        1 & 0 & 0\\
        0 & r^2 \sin^2 \xi & 0\\
        0 & 0 & r^2\\
    \end{bmatrix},
\end{equation}
We choose $f$ to be of the form, 
\begin{equation}
    f=\frac{L}{1+\exp -kR}
    \label{eq:logistic}
\end{equation}
Here, $L$ and $k$ are parameters controlling the point defect distribution.  We
may mention that the equation above is a choice and other choices are possible
permitting us to model different kinds of point defects like extra matter and
vacancy. Note that Eq. \ref{eq:logistic} depends only on the radial coordinate.
We assume that no displacement or traction conditions are applied to the body.
Because of the spherical symmetry, the equilibrium configuration is also assumed
to be radially symmetric. This leads to the following expressions for the
deformation gradient and right Cauchy-Green deformation tensor, 
\begin{equation}
    F_i^I=\begin{bmatrix}
        \frac{\partial r}{\partial R} & 0 & 0\\
        0 & 1 &0\\
        0 & 0 & 1\\
    \end{bmatrix};\;\;\;
    C_{IJ}=\begin{bmatrix}
        \left(\frac{\partial r}{\partial R} \right)^2 & 0 & 0\\
        0 & r^2 & 0\\
        0 &  0 & r^2\\
    \end{bmatrix}.
\end{equation}
The trace of $\tb{C}$, $\tb{C}^{-1}$ and the Jacobian of deformation are given
by,
\begin{equation}
    G^{IJ}C_{IJ}= \left(\frac{\partial r}{\partial R}
    \right)^2+2\left(\frac{r}{R}\right)^2;\;\;\;
    G_{IJ}(C^{-1})^{IJ}= \left(\frac{\partial r}{\partial R} \right)^{-2}
    +2\left(\frac{r}{R}\right)^{-2};\;\;\;
    %C^{IJ}C_{IJ}=\left(\frac{\partial r}{\partial R}
    %\right)^4+\left(\frac{r}{R}\right)^4;\;\;\;
    J=\frac{\partial r}{\partial R}\left(\frac{r}{R}\right)^2
\end{equation}
The equilibrium configuration for the assumed Weyl connection is obtained by
solving the Euler-Lagrange equation for the stored energy; these equations are
nothing but the equilibrium equations given in Eq.\ref{eq:mechEquilibrium}.  We
use a finite element based  numerical procedure to minimize the stored energy
which is discussed in Appendix-\ref{appendix}. The radial component of the Weyl
one-form and the radial displacement at equilibrium are shown in Fig.
\ref{fig:oneForm_Displacement}. Fig.\ref{fig:radialHoopstress} compares the
radial and hoop stresses computed using linear elasticity as well as the present
methodology.  The linear elastic solutions were obtained by Teodosiu
\cite{teodosiu2013}. The stresses in the linear elastic case was predicted by
assuming the elastic constants of the inclusion and the matrix to be same; the
values of $\lambda$ and $\mu$ were assumed to be 1 and $0.3$ respectively. The
difference in volume between the inclusion and the cavity was taken as -0.005.
The same elastic constants are assumed for our model as well. The constant $L$
and $k$ were in Eq. \ref{eq:logistic} are assumed to be 1.35 and 17.5
respectively.  The radial and hoop stress stress distributions for the matrix
and inclusion in the linear elastic case are given by \cite{teodosiu2013},
\begin{equation}
    \sigma^r_r =  
    \Bigg \{
        \begin{array}{ll}
            -\frac{4\mu C}{r_0^3};  &  0\leq r \leq r_0 \\
            -\frac{4 \mu C}{r^3}\left(1+\frac{r^3}{R_0^3}\right); &  r_0 \leq r
            \leq R_0
        \end{array}
\end{equation}

\begin{equation}
    \sigma^\theta_\theta =  
    \Bigg \{
        \begin{array}{ll}
            -\frac{4\mu C}{r_0^3};  &  0\leq r \leq r_0 \\
            \frac{2\mu C}{r^3}\left(1+2\frac{r^3}{R_0^3}\right); &  r_0 \leq r
            \leq R_0
        \end{array}
\end{equation}
where, $r_0$ is the radius of the cavity into which the inclusion is forced,
$R_0$ the radius of the sphere, $C = -\frac{v'}{4\pi\left(1+\frac{4
\mu}{3K'}\right)}$ and $v'$ the difference in volume between the inclusion and
the cavity. From Fig \ref{fig:radialHoopstress}, it is seen that the radial
stress predicted by the present method compares well with the linear elastic
approach, even though the former produces a milder gradient. This may be owing
to a diffused representation of the interface between the inclusion and the
matrix.  The diffused representation manifests in the hoop stress as well; the
unphysical discontinuity in the hoop stress via linear elasticity is absent in
our predictions.  Higher hoop stresses are predicted by our method, which is
essentially due to the smoothness enforced by the diffused representation of the
shrink-fit interface. The diffused representation has also the important
advantage that interfaces need not be tracked explicitly. 

%where, $C_1=\frac{4 \mu C}{3KR^3}$, $C'=\frac{4 \mu
%C}{3K'}\left(\frac{1}{R^3}-\frac{1}{r_0} \right)$ and $C_1=\frac{v'}{4\pi}\left(
%1+\frac{4\pi r^3_0}{3KR^3}+\frac{4 \mu}{3 K'}\left(1-\frac{r_0^3}{R^3} \right)\right)^{-1}$

\section{Stress-free configurations}
We now discuss a framework to calculate non-trivial configurations which are in
a stress-free state in the presence of metrical defects. This problem is
technologically important, since a manufacturing process may, in principle, be
tuned to optimize the defect distribution such that during operation it is close
to a state of zero stress.  Arteries are a good example of systems which are
known to optimize their stress levels by accumulating extra-matter in the form
of growth.  In the context of zero stress configurations, we ask if, for a given
deformation, it is possible to put the body in a state of zero stress by a
distribution of metrical defects. Using our thermo-mechanical theory, we obtain
a condition for the one-form such that body is in a state of zero stress.
Applying this condition, we compute non-trivial configurations which are in a
state of zero stress for a given temperature change. For the sake of analytical
tractability, we assume the body to be in a state of thermal equilibrium. The
relaxation experienced by the defects to reach the equilibrium state is also
ignored. For the calculations performed in this section, we adopt the
constitutive rule discussed in Section \ref{sec:specificConstitution}.

We now deduce the constraint imposed on the one-form by the condition of zero
stress; this condition is obtained by setting the second Piola stress to zero.
Even though any stress measure may be utilized to obtain the zero stress
condition, expressing it in terms of the second Piola stress is more convenient
as the free energy is postulated in the reference configuration. For the assumed
free energy, the zero stress condition is given by, 
\begin{equation}
    \frac{\mu}{2}G^{IJ}+\frac{1}{2}\left(-\mu+\lambda\ln(\bar{J})
    \right)(\bar{C}^{-1})^{IJ}+k_1\bar{C}^{IJ}{\phi'}^K
    {\phi'}_K=0.
    \label{eq:stressFreeCondition}
\end{equation}
It should be noted that defects influence the stresses only through
$\phi^{'I}\phi'_{I}$, which is nothing but the magnitude of the one-form given
by $\tb{G}^{-1}(\phi',\phi')$. Contracting the last equation with the reference
metric tensor, we arrive at the following equation relating $\phi^{'I}\phi'_{I}$
and the components of the right Cauchy-Green deformation tensor,
\begin{equation}
    \phi^{'I}\phi'_{I}=\frac{k_1}{\bar{C}^{IJ}\bar{G}_{IJ}}\left[
            \frac{3 \mu}{2}+\frac{1}{2}\left( -\mu+ \lambda
            \ln(\bar{J}) \right)(\bar{C}^{-1})^{IJ}\bar{G}_{IJ}
            \right].
\end{equation}
If the deformation is assumed to be known, then the equation above may be
thought of as a constraint on the Weyl one-form. The equilibrium distribution of
the Weyl one-form may then be arrived at based on the minimization problem for
$\phi$ given in Eq. \ref{eq:minimumOneform} with the last equation acting as a
constraint.

On the other hand, if one assumes the defect density to be given and ask if
there exists a deformation which renders the body stress free, Eq.
\ref{eq:stressFreeCondition} becomes a condition on the deformation gradient.
However, the deformation gradient computed from the Eq.
\ref{eq:stressFreeCondition} need not be integrable,  i.e. there may not exist a
deformation map whose derivative is the $\tb{F}$ computed from Eq.
\ref{eq:stressFreeCondition}. Integrability of $\tb{F}$ is guaranteed, if the
anholonomy associated with the tangent vector in the range of $\tb{F}$ vanishes.
If we denote the tangent vector in the range of $\tb{F}$ by
$\{\tb{e}_1,\tb{e}_2,\tb{e}_3\}$, then
 \begin{equation}
     \tb{e}_i=F_i^I \frac{\partial}{\partial X^I}.
 \end{equation}
We are asking for the condition under which $\tb{e}_i=\frac{\partial }{\partial
x^i}$ for some coordinates $x^i$. The anholonomy associated with the vector
fields $\tb{e}_i$ is given by,
\begin{equation}
    [\tb{e}_i, \tb{e}_j]=A_{ij}^k\tb{e}_k 
\end{equation}
$[.,.]$ denotes the Lie bracket between the vector fields. If all the
Lie brackets vanish, then one can find a coordinate system such that the each
vector of the frame field can be described as a tangent to some coordinate line.
In terms of the deformation gradient, the components of anholonomy may be
found as,
\begin{equation}
        A_{ij}^k=\left(F_i^J\frac{\partial F_j^k}{\partial X^J}
        -F_j^J\frac{\partial F_i^k}{\partial X^J}\right).
\end{equation}
The condition for integrability of $\tb{F}$ may now be written as,
\begin{equation}
    A_{ij}^k=0
    \label{eq:integrablityOfF}
\end{equation}
For multiply connected bodies, additional conditions are required depending on
the muti-connectedness of the domain. For the integrability of the deformation
gradient in such a domain, see \cite{yavari2013}. In the following, we explore
simple distributions of defects which keep the body in a stress free state.

\subsection{Stress free configuration under temperature change}
We specialize the zero stress equations obtained in the previous sub-section to
incompatible strains caused by temperature change.  Assuming the temperature
field to be given, the variable of Weyl transformation $s$ is determined using
Eq. \ref{eq:thermalExpansion}.  It is well established that not all temperature
distributions can be made stress-free by applying deformation \cite{boley2012}.
The necessary condition for a body with temperature change to be stress free is
Eq.  \ref{eq:stressFreeCondition}, with the one-form given by $\phi=\text{d}s$.
Substituting the condition above into Eq. \ref{eq:stressFreeCondition} leads to,
\begin{equation}
    \frac{\mu}{2}G^{IJ}+\frac{1}{2}(-\mu+ \lambda \log(\bar{J}))(\bar{C}^{-1})^{IJ}
    + k_1\bar{C}^{IJ}G^{KL}s_{,L}
    s_{,K}=0.
    \label{eq:zeroStressConditionTemperature}
\end{equation}
The last equation is an implicit relation for the deformation gradient. In
addition, the integrability of $\tb{F}$ discussed in Eq.
\ref{eq:integrablityOfF} also needs to be imposed.  

Now, we compute the deformed shape of the body with a constant temperature field
imposed, i.e. $s$ is a constant in Eq. \ref{eq:stressFreeCondition}. Assuming a
Cartesian coordinate system for the reference and deformed configurations, the
metric tensors in these configurations may be written as $\delta_{IJ}$ and
$e^s\delta_{ij}$ respectively. From Eq.  \ref{eq:zeroStressConditionTemperature},
we arrive at the following relation for the deformation gradient, 
\begin{equation}
    \tb{F}^T \tb{F}=\frac{1}{\mu e^{2s}}\left(\mu-
    \frac{3}{2}\lambda s-\lambda \log(\det(\tb{F}))\right)\tb{I}.
    \label{eq:stressFreeConditionTemp}
\end{equation}

\begin{figure}
    \centering
    \includegraphics[width=.45\linewidth]{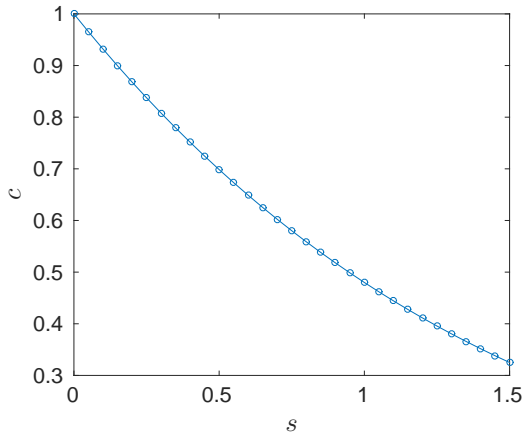}
    \caption{Value of $c$ for different values of $s$; the material constants
    $\lambda$ and $\mu$ are both taken as 1.}
    \label{fig:dilatationVsf}
\end{figure}

We assume a solution of the form $F=c \bs{\Lambda}$, $\bs{\Lambda}$ $\in$ SO(3)
and $c\in \mathbb{R}$. Substituting these in Eq.  \ref{eq:stressFreeCondition},
we arrive at the following equation for $c$,
\begin{equation}
    c^2=\frac{1}{\mu e^{2s}}\left(\mu- \frac{3\lambda s}{2}-3\lambda\log c \right)
    \label{eq:constantDilation}
\end{equation}
The SO(3) part is irrelevant since we are looking at a body with no displacement
and traction conditions applied. If we set $s=0$ in the above equation, it
reduces to $c^2 + \frac{3 \lambda}{\mu} \log c =1$. The solution to this
equation is $c=1$. This solution recovers the reference configuration of the
body.  Non-trivial stress-free configurations for different values of $s$ may be
computed through a numerical root finding technique, applied to Eq.
\ref{eq:constantDilation}. The values of $c$ computed for different values $s$
are shown in Fig. \ref{fig:dilatationVsf}.
The curve shown in Fig.  \ref{fig:dilatationVsf} is slightly nonlinear; this is
because of the nonlinear constitutive rule assumed for stress. Having computed
the deformation gradient, one needs to establish that $\tb{F}$ is compatible.
For the present scenario, compatibility is not an issue as the deformation
gradient is constant. The above calculation also establishes the classical
result that, for an isotropic body subjected to constant temperature field and
free-free boundary conditions, the deformation is volumetric and without stress.
The constitutive relation for stress also plays a major role in determining
stress free configurations. It may so happen that certain constitutive rules
would not permit any solution for a stress free configuration and constructing
such constitutive rules should be an interesting exercise. The role played by
constitutive inequalities is also worth exploration.

\section{Conclusion}
We have reported a geometric theory for thermo-mechanical deformation of bodies
with metrical defects. We have exploited the geometric setting initially
proposed by Weyl in the context of general relativity to represent bodies with
metrical defects. These defects are identified with the incompatibility of the
connection and the metric. Our proposal on the thermo-mechanics of
defect-mediated deformation has completely dispensed with the notion of an
intermediate configuration. The defect equilibrium, described by the Weyl
one-form, is obtained as a critical point of the free energy. The Weyl
transformation is used to model the interaction of incompatibilities introduced
by temperature change.  To include dissipation in the defect evolution, we
introduced a viscous term in the defect equilibrium equation, which made the
evolution of one-form a gradient flow.  Using the laws of thermodynamics,
restrictions have been obtained on the constitutive rules. The important problem
of relating stress and strains in a non-Euclidean setting is resolved via the
Doyle-Ericksen formula. 

We have applied our theory to a diffused version of the shrink-fit problem which
represents point defects. With the equilibrium configurations computed using a
minimization procedure, we have shown that the shifted critical point of the
stored energy in the presence of a Weyl connection results in residual stresses.
This solution could also recover aspects of the linear elastic shrink-fit
problem.  Other than mathematical expedience, a diffused representation is also
practically meaningful, say in the context of additive manufacturing processes
that materials in the form of thin layers.  In addition, we have derived
conditions for a defective body to be in a state of zero stress.  Using this
condition, we have been able to recover zero stress configurations resulting
from simple temperature distribution. Also recovered in the process are the well
known linear elastic results, albeit in a nonlinear setting. 

We have focused mostly on the geometric aspects in the model and addressed only
a few simple boundary value problems. We have left unaddressed the
characterization of the Weyl one-form to a particular kind of point defect. This
characterization can be done using molecular dynamic calculations or from
density function theory calculations.  Such characterizations do exist for the
dipole tensor used in the linear elastic case. A more interesting problem would
be to study the interaction of dislocations and point defects within a similarly
geometric framework. This would require the connection to have torsion. The
numerical solution procedure presently adopted is somewhat simplistic and does
not respect all the geometric features of the model. Geometrically motivated
discretization schemes based on finite element exterior calculus might be a very
good approach to arrive at efficient and accurate numerical solution schemes.

\appendix
\section{Numerical solution procedure}
\label{appendix}
This section discusses the numerical procedure used to obtain the solution of
the minimization problem discusses in Section \ref{sec:pointDefect}. We
formulate the (mechanical) equilibrium of a body with a point defect as a
minimization problem over a Weyl manifold, with the Weyl connection specified.
As discussed earlier, the connection on the manifold modifies the critical
points of the stored energy function. These configurations are computed using a
finite element based discretization procedure. Since we are looking at a
radially symmetric problem, only the radial displacement is discretized using a
three noded quadratic Lagrange finite element. The minima of the discretized
stored energy function is computed using Newtons' method which requires the
first and the second derivatives of the discrete stored energy. The first
derivative of the stored energy is often called the residual force, we denote it
by $\tb{R}$. The condition $\tb{R}=\tb{0}$ represent the equilibrium of forces
at the node. The discrete equilibrium is analogous to the linear momentum
equation given in Eq.  \ref{eq:mechEquilibrium}. Since the body has non-trivial
connection, we use following equation to compute an incremental change in the
deformation.
\begin{equation}
    F^a_{A}=H^a_A+\left(\gamma^a_{bc}H^b_A u^c+\frac{\partial
    u^a}{\partial X^A}\right)
    \label{eq:perturbedDefGrad}
\end{equation}
In the above equation, $\tb{u}$ is an incremental displacement field
superimposed on an deformation $\varphi_0$ with deformation gradient $\tb{H}$ and
$\gamma_{jk}^i$ are the connection coefficients of the deformed configuration. The second term in Eq. \ref{eq:perturbedDefGrad} accounts for non-trivial
connection associated with the configuration $\varphi_0$, the proof of the above
equation can be found in \cite{mfe}.
The components of the connections in spherical coordinate system is given as,
\begin{equation}
    \nonumber
    \gamma^r=
    \begin{bmatrix}
        -\frac{1}{2}\partial_r f & -\frac{1}{2}\partial_\theta f &
        -\frac{1}{2}\partial_\xi f\\
        -\frac{1}{2}\partial_\theta f & -r\sin \xi & 0 \\
        -\frac{1}{2}\partial_\xi f & 0 & -r
    \end{bmatrix};
    \gamma^\theta =
    \begin{bmatrix}
        0 & \frac{1}{r}-\frac{1}{2}\partial_r f & 0\\
        \frac{1}{r}-\frac{1}{2}\partial_r f & -\frac{1}{2}\partial_\theta f &
        \cot \xi -\frac{1}{2}\partial_\xi f\\
        0 & \cot \xi -\frac{1}{2}\partial_\xi f& 0
    \end{bmatrix};
\end{equation}
\begin{equation}
    \gamma^\xi=
    \begin{bmatrix}
        0 & 0 & \frac{1}{r}-\frac{1}{2}\partial_r f\\
        0 & -\sin \xi \cos \xi & -\frac{1}{2}\partial_\theta f\\
        \frac{1}{r}-\frac{1}{2}\partial_r f & -\frac{1}{2}\partial_\theta f &
        -\frac{1}{2}\partial_\xi f
    \end{bmatrix}
\end{equation}
The residual force can then be computed
using the relation, 
\begin{equation}
    \tb{R}=\frac{d}{d \epsilon}|_{\epsilon=0}W(\epsilon)
\end{equation}
Where, $W(\epsilon)$ denotes the perturbed stored energy about the configuration
$\varphi_0$, which is obtained using the relation $\varphi_0+\epsilon\tb{u}$,
here $\epsilon$ is a small parameter. The perturbed deformation gradient is
obtained using the relation Eq.  \ref{eq:perturbedDefGrad}. The finite element
approximation for the incremental radial displacement $u_r$ is denoted by
$u^h_r$ and it is given by,
\begin{equation}
    u_r^h=\sum_{i=1}^{i=N} N_i u_r^i
\end{equation}
In the above equation $u_r^i$ and $N_i$ denotes the incremental radial
displacement and shape function at the $i^{th}$ node. The incremental radial
displacement and the nodal shape function at all nodes are denoted by $\tb{u}^h$
and $\bs{\Upsilon}$ respectively.  In terms of the nodal displacement and stored
energy function, the condition for discrete mechanical equilibrium can be
written as,
\begin{equation}
    \frac{\partial W}{\partial \tb{u}^h}=\tb{0}
\end{equation}
The above equation is nothing but the necessary condition of an extrema for a
finite dimensional extremization problem.  For the assumed stored energy
function, the discrete equilibrium equation is given by,
\begin{equation}
    \frac{\mu}{2}\frac{\partial I_1}{\partial \tb{u}^h}+\frac{1}{J}(\lambda
    \log(J)-\mu)\frac{\partial J}{\partial \tb{u}^h}=0
\end{equation}
The variables $\frac{\partial I_1}{\partial \tb{u}^h}$ and $\frac{\partial
J}{\partial \tb{u}^h}$ denotes the directional derivatives computed using the
perturbation given in Eq. \ref{eq:perturbedDefGrad}, whose expressions are given
as,
\begin{eqnarray}
    \frac{\partial I_1}{ \partial
    \tb{u}}=&2\left(\frac{dr}{dR}-\frac{u_r}{2}\frac{df}{dR}
    \frac{dr}{dR}+\frac{du_r}{dR}\right)
    \left(-\frac{1}{2}\frac{df}{dR}\frac{dr}{dR}\bs{\Upsilon}+\frac{d \bs{\Upsilon}}{dR}
    \right)+\frac{4r^2}{R^2}\left(1+u_r\left(\frac{1}{r}-\frac{1}{2}\frac{df}{dR}\right)
    \right)\left(\frac{1}{r}-\frac{1}{2}\frac{df}{dR}\right)\bs{\Upsilon}\\
    \nonumber
    \frac{\partial J}{\partial
    \tb{u}}=&\left(\left(1+u_r\left(\frac{1}{r}-\frac{1}{2}\frac{df}{dR}\right)\right)^2
    \left(-\frac{1}{2}\frac{df}{dR}\frac{dr}{dR}\bs{\Upsilon}+\frac{d
    \bs{\Upsilon}}{dR}\right)\right) \left(\frac{r}{R}\right)^2
    +2\left(1+u_r\left(\frac{1}{r}-\frac{1}{2}\frac{df}{dR}\right)\right)\\
    &\left(\frac{dr}{dR}-\frac{u_r}{2}\frac{dr}{dR}\frac{df}{dR}+\frac{du_r}{dr}\right)
    \left(\frac{1}{r}-\frac{1}{2}\frac{df}{dR}\right)\left(\frac{r}{R}\right)^2
 \bs{\Upsilon}
\end{eqnarray}
In the above equation, $u_r$ denotes the displacement computed from the last
iteration and $r$ denotes the last converged radius and $\frac{d
\bs{\Upsilon}}{dR}$ denotes the derivative of the nodal basis with respect to the
radial co-ordinates. The Hessian (tangent stiffness matrix) required for
Newtons' method for the assumed stored energy function is given by,
\begin{equation}
    \frac{\partial^2 W}{\partial \tb{u} \partial
    \tb{u}}=\frac{\mu}{2}\frac{\partial ^2 I_1}{\partial \tb{u} \partial
    \tb{u}}+\frac{1}{J^2}(\lambda+\mu-\lambda \log J)\frac{\partial J}{\partial
    \tb{u}} \otimes \frac{\partial J}{\partial \tb{u}} 
\end{equation}
The second derivatives of $I_1$ and $J$ are computed as,
\begin{equation}
    \frac{\partial I_1}{\partial \tb{u} \partial
    \tb{u}}=2\left(-\frac{1}{2}\frac{dr}{dR}\bs{\Upsilon}+\frac{d
    \bs{\Upsilon}}{dR}
    \right)\otimes\left(-\frac{1}{2}\frac{dr}{R}\frac{df}{dr}\bs{\Upsilon}+\frac{d
    \bs{\Upsilon}}{dR}
    \right)+4\frac{r^2}{R^2}\left(\frac{1}{r}-\frac{1}{2}\frac{df}{dR}\right)^2\bs{\Upsilon}\otimes
    \bs{\Upsilon}
\end{equation}
\begin{eqnarray}
    \nonumber
    \frac{\partial J}{\partial \tb{u} \partial \tb{u}}
    =
    2\left(1+u_r\left(\frac{1}{r}-\frac{1}{2}\frac{df}{dR}\right)\right)\left(\frac{1}{r}-\frac{1}{2}\frac{df}{dR}\right)
    \left(\bs{\Upsilon}\otimes\left(-\frac{1}{2}\frac{dr}{dR}\bs{\Upsilon}+\frac{d
    \bs{\Upsilon}}{dR}\right)+\left(-\frac{1}{2}\frac{dr}{dR}\bs{\Upsilon}+\frac{d
    \bs{\Upsilon}}{dR}\right)\otimes \bs{\Upsilon} \right)\\
    +2\left(\frac{1}{r}-\frac{1}{2}\frac{df}{dR}\right)^2\left(\frac{dr}{dR}-\frac{u_r}{2}\frac{dr}{dR}\frac{df}{dR}+\frac{du_r}{dR}
    \right)\left(\frac{r^2}{R^2}\right)\bs{\Upsilon}\otimes\bs{\Upsilon}
\end{eqnarray}

\newpage
\bibliographystyle{abbrv}
\bibliography{main}
\end{document}